\documentclass[prl,twocolumn,superscriptaddress,floats]{revtex4}

\usepackage[dvips]{graphicx}
\usepackage{amsmath}
\usepackage{booktabs}
\usepackage{dcolumn}
\usepackage{sidecap}

\usepackage{rotating}

\newcolumntype{d}{D{.}{.}{-1}}

\newcommand{\tph}{90 K}
\newcommand{\cacr}{CaCrO$_3$}
\newcommand{\cacro}{CaCrO$_3$}

\begin{document}

\textheight 24.5 true cm

\title{CaCrO$_3$: an anomalous antiferromagnetic metallic oxide}

\author{A. C. Komarek }
\affiliation{II. Physikalisches Institut, Universit\"{a}t zu
K\"{o}ln, Z\"{u}lpicher Str. 77, D-50937 K\"{o}ln, Germany}

\author{S.V.~Streltsov}
\affiliation{Institute of Metal Physics, S.Kovalevskoy St. 18,
620041 Ekaterinburg GSP-170, Russia}

\author{M. Isobe }
\affiliation{Institute for Solid State Physics, The University of
Tokyo, 5-1-5 Kashiwanoha, Kashiwa, Chiba 277-8581, Japan}

\author{T. M\"{o}ller}
\affiliation{II. Physikalisches Institut, Universit\"{a}t zu
K\"{o}ln, Z\"{u}lpicher Str. 77, D-50937 K\"{o}ln, Germany}

\author{M. Hoelzel}
\author{A.~Senyshyn}
\affiliation{Technische Universit\"{a}t Darmstadt, Material und
Geowissenschaften, Petersenstrasse 23, D-64287 Darmstadt, Germany
und Technische Universit\"{a}t M\"{u}nchen, FRM-II,
Lichtenbergstr. 1, D-85747 Garching, Germany}

\author{D. Trots}
\affiliation{Hasylab/DESY, Notkestr. 85, D-22607, Hamburg,
Germany}

\author{M.T. Fern\'andez-D\'iaz}
\affiliation{Institut Laue-Langevin, 38042 Grenoble, France}

\author{T. Hansen}
\affiliation{Institut Laue-Langevin, 38042 Grenoble, France}

\author{H. Gotou}
\affiliation{Institute for Solid State Physics, The University of
Tokyo, 5-1-5 Kashiwanoha, Kashiwa, Chiba 277-8581, Japan}

\author{T. Yagi}
\affiliation{Institute for Solid State Physics, The University of
Tokyo, 5-1-5 Kashiwanoha, Kashiwa, Chiba 277-8581, Japan}

\author{Y. Ueda}
\affiliation{Institute for Solid State Physics, The University of
Tokyo, 5-1-5 Kashiwanoha, Kashiwa, Chiba 277-8581, Japan}

\author{V.I.~Anisimov}
\affiliation{Institute of Metal Physics, S.Kovalevskoy St. 18,
620041 Ekaterinburg GSP-170, Russia}

\author{M. Gr\"{u}ninger}
\affiliation{II. Physikalisches Institut, Universit\"{a}t zu
K\"{o}ln, Z\"{u}lpicher Str. 77, D-50937 K\"{o}ln, Germany}

\author{D.I.~Khomskii}
\affiliation{II. Physikalisches Institut, Universit\"{a}t zu
K\"{o}ln, Z\"{u}lpicher Str. 77, D-50937 K\"{o}ln, Germany}

\author{M. Braden}
\affiliation{II. Physikalisches Institut, Universit\"{a}t zu
K\"{o}ln, Z\"{u}lpicher Str. 77, D-50937 K\"{o}ln, Germany}

\date{\today}

\pacs{PACS numbers:}


\begin{abstract}

Combining infrared reflectivity, transport, susceptibility and
several diffraction techniques, we find compelling evidence that
CaCrO$_3$ is a rare case of a metallic and antiferromagnetic
transition-metal oxide with a three-dimensional electronic
structure. LSDA calculations correctly describe the metallic
behavior as well as the anisotropic magnetic ordering pattern of
$C$ type: The high Cr valence state induces via sizeable $pd$
hybridization remarkably strong next-nearest neighbor
interactions stabilizing this ordering. The subtle balance of
magnetic interactions gives rise to magneto-elastic coupling,
explaining pronounced structural anomalies observed at the
magnetic ordering transition.
\end{abstract}

\maketitle

Strongly correlated electron systems including the wide class of
transition-metal oxides  exhibit a quite general relation between
magnetic order and electrical conductivity \cite{Goodenough}:
ferromagnetism typically coexists with metallic conductivity,
whereas insulators usually exhibit antiferromagnetism. It is
always a challenge to understand exceptions from this rule. The
rare observations of ferromagnetism in insulating transition-metal
oxides most often are due to a particular type of orbital ordering
\cite{Khomskii-97}. The few examples of antiferromagnetic (AFM)
metals, e.g., (La/Sr)$_3$Mn$_2$O$_7$ \cite{327-mang} or
Ca$_3$Ru$_2$O$_7$ \cite{327-ruth}, are characterized by reduced
electronic and structural dimensionality, and the
antiferromagnetic order corresponds to a stacking of
ferromagnetic (FM) layers. Here we report the discovery of a
three-dimensional transition-metal oxide with metallic
conductivity, antiferromagnetic exchange interactions, and
$C$-type antiferromagnetic order: the perovskite CaCrO$_3$.


Perovskites containing Cr$^{4+}$ (CaCrO$_3$, SrCrO$_3$, and PbCrO$_3$) were already studied
previously~\cite{Zhou-06,Williams-06,Chamberland-67,goodenough,weiher,ortega},
but neither the details of the crystal structure nor the nature of the magnetic ordering are
known. Only very recently evidence for
C-type AFM order was reported in multi-phase samples of SrCrO$_3$
\cite{ortega}.
Regarding the conductivity, the existing data are controversial.
In Refs.~\cite{Chamberland-67,weiher}
CaCrO$_3$ was claimed to be metallic,
but more recently insulating behavior has been reported
\cite{Zhou-06}. A similar controversy persists also for SrCrO$_3$,
which should definitely be more metallic than CaCrO$_3$ due to the
less distorted crystal structure, but metallic behavior was
observed in Ref.\ \cite{Zhou-06} only under pressure. These
controversies most likely are connected with  the difficulty to
prepare high-quality stoichiometric materials and with the lack of
large single crystals.

\cacr\ exhibits an orthorhombic GdFeO$_3$-type perovskite
structure and early magnetization measurements indicate a
magnetic transition at 90\ K \cite{goodenough}, which is
confirmed in our samples. Two electrons occupy the Cr $3d$ shell
(S=1), rendering the material electronically similar to
insulating $R$VO$_3$ \cite{q1} (also $3d^2$) and to metallic
(Ca/Sr)RuO$_3$ ($4d^2$) \cite{ruthenates}. CaCrO$_3$ shows an
unusually high transition-metal valence,
 Cr$^{4+}$, which may lead to a small or even negative charge-transfer gap
\cite{ZSA,Khomskii-97-2}, i.e., holes in the O band. In CrO$_2$
with rutile structure and edge-sharing CrO$_6$ octahedra, the
negative charge-transfer gap leads to
self-doping~\cite{Korotin-98} and to the appearance of a
ferromagnetic metallic state. In contrast, the layered perovskite
Sr$_2$CrO$_4$ with corner-sharing octahedra and $\sim$
180$^{\circ}$ Cr-O-Cr bonds is an AFM Mott-Hubbard insulator with
a gap of about 0.2 eV \cite{matsuno05}.

Combining diffraction, macroscopic and infrared reflectivity
measurements with LSDA as well as with LSDA+U calculations we
have studied the properties of \cacro . We find that \cacro \ is
an antiferromagnetic metallic transition-metal oxide with a
$C$-type magnetic structure. According to LSDA calculations, the
magnetic order arises from competing nearest-neighbor (nn) and
next-nearest-neighbor (nnn, "diagonal") exchange interactions,
which result from a sizeable $pd$ hybridization. Remarkably, the
magnetic transition at T=\tph\ causes pronounced anomalies in
structural and transport properties.

Polycrystalline CaCrO$_3$ was prepared by a solid state reaction
of CaO and CrO$_2$ under 4\ GPa at 1000$^o$C for 30 minutes. The
obtained samples of stoichiometric reactions always include a
varying amount of the impurities of Cr$_2$O$_3$ and CaCr$_2$O$_4$.
This impurity problem has also been reported by Goodenough et al.
\cite{goodenough}. A small excess of CaO (5-10$\%$), however,
almost completely eliminated these impurities, and the excess CaO
could be washed out with distilled water.  Close inspection showed
that single-crystalline grains of up to 0.1mm diameter were
obtained by this procedure as well. Powder neutron measurements
were performed on the SPODI diffractometer at the FRM-II reactor
($\lambda = 1.548 $\AA) as well as on the D20 high-flux
diffractometer at the ILL ($\lambda = 2.4233 $\AA). Lattice
parameters have been studied with synchrotron radiation at the
beamline B2 at Hasylab/DESY ($\lambda = 0.75 $\AA) using an image
plate detector for temperatures between 15\ K and 1063\ K.\ At
about 710\ K a starting sample decomposition, however, allowed us
to obtain reliable data only up to $\sim$800\ K. X-ray
single-crystal structure analysis was performed on a Bruker
X8-Apex diffractometer using Mo-K$_\alpha$ radiation between 90\ K
and 300 K.\@ Although the sample showed a superposition of six
different twin-domain orientations, a satisfactory intensity
integration was achieved due to the low splitting of the
pseudo-cubic parameters. Final R-values referring to the
intensities were between 2 and 4\%. This experiment confirms the
close to perfect stoichiometry of
our samples. 
The electrical resistivity $\rho(T)$ was measured by an AC
four-point method on a pellet of \cacr \ powder which was
cold-pressed at 12.5 kbar. The infrared reflectivity $R(\omega)$
of a cold-pressed pellet was determined between 7 meV and 0.9 eV
using a Bruker IFS 66v/S Fourier-transform spectrometer. For the
reference measurement we used {\em in-situ} Au evaporation. The
real part $\sigma_1(\omega)$ of the optical conductivity has been
obtained via a Kramers-Kronig analysis, for which $R(\omega)$ has
been extrapolated to lower and higher frequencies using a
conventional Drude-Lorentz fit.

The lattice parameters of \cacro\ determined by
synchrotron-radiation powder diffraction are shown in Fig.\ 1.
All three orthorhombic parameters exhibit a step-like anomaly at
the  magnetic ordering temperature determined by the SQUID
susceptibility and neutron diffraction measurements, $T_N$=\tph .
Although the sudden changes in the lattice constants are rather
strong, up to 0.5\% for $c$, there is no visible effect in the
lattice volume. Whereas $c$ shrinks, $a$ and $b$ elongate upon
cooling, yielding a flattening of the $Pbnm$ structure. We
emphasize that there is no evidence for phase mixture apart very
close to $T_N$. Close inspection of the temperature dependence
suggests that this lattice flattening already starts at much
higher temperatures.

\begin{figure}[t]
\begin{center}
\includegraphics*[width=0.96\columnwidth]{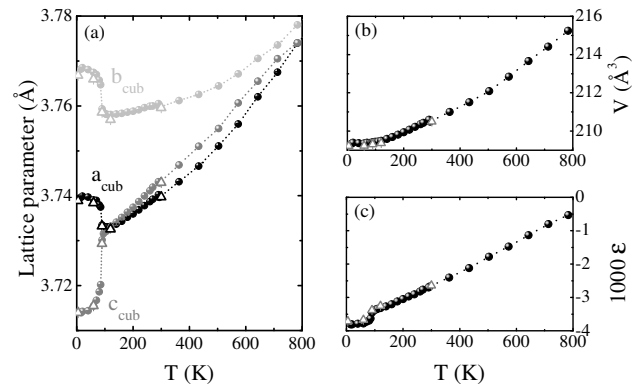}
\end{center}
\caption{a) Orthorhombic lattice parameters $a$, $b$, and $c$,
scaled to the parameter of a cubic perovskite:
$a_{cub}=a/\sqrt{2}$, $b_{cub}=b/\sqrt{2}$ and $c_{cub}=c/2$. b)
and c) show the orthorhombic splitting $\varepsilon$ and the
lattice volume $V$. Circles refer to synchrotron and triangles to
neutron diffraction results. }
\end{figure}

The neutron powder and the x-ray single-crystal experiments give
the full structural information including the positional and the
displacement parameters, from which all bond angles and distances
can be calculated. The results are given in Fig.\ 2. The
GdFeO$_3$-type structure (space group $Pbnm$) develops out of the
ideal perovskite structure by rotating, angle $\Phi$, and tilting,
angle $\Theta$, the CrO$_6$ octahedra \cite{komarek}. Between 3.5
and 300\ K, $\Theta =10.5^\circ$ and $\Phi = 8.2^\circ$ are nearly
constant, reflecting a sizeable structural distortion. The
combination of tilt and rotation yields two distinct O positions:
apical O1 out-of-plane and O2 in the $ab$ plane. Regarding a
distortion of the basal plane of the octahedron, we do not find a
splitting in the Cr-O2 distances but a weak temperature
independent elongation of the octahedron parallel to $a$, i.e.\
the O2-O2 edges are different (see Fig.\ 2). In addition, we find
an overall flattening of the octahedron following the flattening
of the lattice at $T_N$: The Cr-O1 (Cr-O2) distance shrinks
(elongates) upon cooling. The compression of the octahedron points
to a temperature-driven redistribution amongst the $t_{2g}$
orbitals increasing the $d_{xy}$ occupation upon cooling into the
magnetically ordered state. In electronically similar
Ca$_2$RuO$_4$, a similar flattening of the octahedron has been
attributed to a pronounced orbital rearrangement
\cite{braden-friedt,liebsch}, but there the effects are about an
order of magnitude larger than in \cacro .

\begin{figure}[t]
\begin{center}
\includegraphics*[width=0.999\columnwidth]{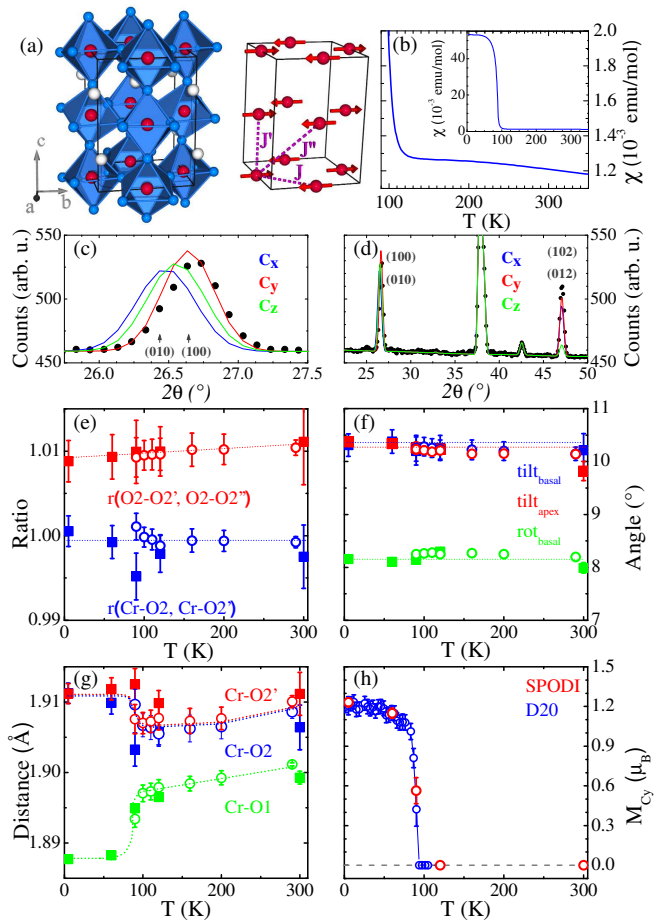}
\end{center}
\caption{(color online)  a) Crystal structure of \cacr\ in space
group \emph{Pbnm} and the C-type magnetic structure of Cr moments
\emph{(red)} indicating the main magnetic interaction paths.  b)
Magnetic susceptibility. c-h) Results of powder neutron
(\emph{squares}) and single crystal X-ray (\emph{circles})
measurements. c): Magnetic (010)/(100) reflection at 3.5 K and
calculated profiles for $C_x$, $C_y$ and $C_z$ type magnetic
order. d): Part of the neutron diffraction pattern at 3.5\ K and
calculated profiles. e): Ratio of O2-O2 and Cr-O2 bond lengths;
f): Octahedral tilt $\Theta$ and rotation $\Phi$ angles. g): Cr-O1
and Cr-O2 bond lengths. h): $C_y$-type ordered magnetic moment in
$\mu_B$. }
\end{figure}

Below $T_N$=\tph\ two strong magnetic peaks emerge at (100) and
(102)/(012) which can unambiguously be attributed to $C_y$-type
AFM order, see Fig.\ 2. Other schemes do not yield the correct
peak positions or fail to describe the intensity ratio. In space
group $Pbnm$ the $C_y$-type order may couple with $F_x$ and $G_z$
components according to the irreducible representation
$\Gamma_{2g}$ \cite{bertaut}. The $F_x$ component perfectly
agrees with the observation of weak ferromagnetism in the
susceptibility, see Fig. 2b). We find a sizeable ordered moment of
1.2$\mu_B$ at low temperature, which, however, is much below the
expected value for a S=1 moment.

The resistivity $\rho(T)$ exhibits $\partial \rho/\partial T < 0$
and a rather small value at 300 K, 0.1$\Omega$m (see inset of
Fig.\ 3). Furthermore, $\rho(T)$ does not diverge towards low $T$
but tends to a finite value. Upon a first cooling cycle we find a
clear jump at $T_N$ most likely due to cracks caused by the
pronounced structural anomalies. A similar jump was observed
close to 90\ K in $\rho(T)$ of a \emph{metallic} single crystal
\cite{weiher}, suggesting that this sample exhibits a fully
comparable magnetic transition and thus can be considered to
represent stoichiometric \cacro . But $\rho(T)$ of
polycristalline CaCrO$_3$ appears to be dominated by grain
boundaries.

\begin{figure}[t]
\begin{center}
\includegraphics*[width=0.8\columnwidth,angle=0]{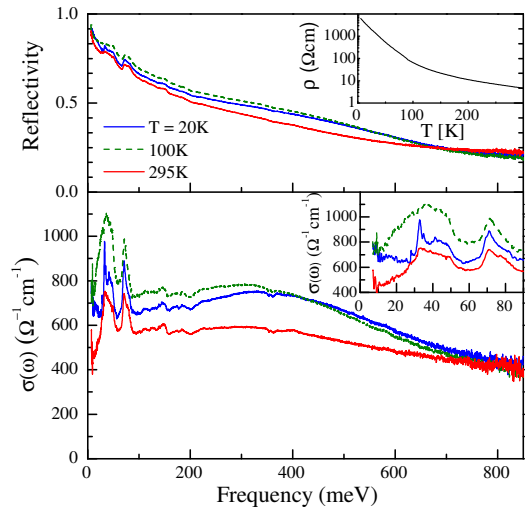}
\end{center}
\caption{Reflectivity (top) and optical conductivity $\sigma_1(\omega)$ (bottom).
The inset in the bottom panel shows $\sigma_1(\omega)$ at low frequencies on an
enlarged scale; the inset in the top panel depicts the resistivity $\rho(T)$.
} \label{figOpt}
\end{figure}

In contrast to DC transport, optical data can reveal the metallic
properties of a polycrystalline metal with insulating grain
boundaries. Figure 3 clearly demonstrates that CaCrO$_3$ is a
metal with a moderate conductivity, $\sigma_1(\omega)$ of the
order of a few hundred to 1000 ($\Omega$cm)$^{-1}$. Typical for a
metal, $R(\omega)$ extrapolates to 1 for $\omega \! \rightarrow \!
0$. Phonons are observed between 20 and 80 meV, and in $R(\omega)$
they are strongly screened by the itinerant charge carriers.
However, the frequency dependence deviates strongly from a typical
Drude behavior. The spectral weight is almost entirely dominated
by a peak at about 350 meV.\@ An increase of $\sigma_1(\omega)$
with decreasing frequency is recovered only below 30 meV at 20
K.\@ Although $\sigma_1(\omega)$ appears to be dominated by
excitations with finite frequency, we emphasize that $R(\omega)$
unambigously demonstrates the presence of free carriers. This
further agrees with the magnetic susceptibility which, above $T_N$
is very small and hardly temperature dependent indicating
itinerant magnetism, see Fig. 2b).

In the case of insulating grain boundaries, $\sigma_1(\omega)$ is
suppressed at low frequencies. However, this can not explain the
peak observed at 350 meV. In our samples of CaCrO$_3$, the typical
grain size is of the order of 20 $\mu$m. Since a wavelength of
$\lambda$ = 20 $\mu$m is equivalent to a photon energy of $\sim$
60 meV, grain-size effects can become important only much below
350 meV \cite{grueninger}. We attribute the peak at 0.35 eV to
excitations from the lower (LHB) to the upper (UHB) Hubbard band.
 Due to the high valence of Cr$^{4+}$, the Cr Hubbard
bands shift down towards the fully occupied O-2p band, whereas
the $pd$ hybridization between Cr and O bands pushes the LHB back
upwards, reducing the effective Coulomb repulsion $U_{\rm eff}$
\cite{goessling} and admixing O-2p states  to the LHB and the UHB
in the same way as it was demonstrated for CrO$_2$
\cite{Korotin-98}.
In insulating Sr$_2$CrO$_4$ with Cr$^{4+}$ in a $d^2$
configuration \cite{matsuno05}, this LHB-UHB excitation was
observed at 1.0 eV. Integrating $\sigma_1(\omega)$ from 7 meV to
0.9 eV in  CaCrO$_3$ yields an effective carrier density $N
\approx 0.1$ per Cr ion. Remarkably, this is very similar to the
spectral weight of the LHB-UHB peak  in Sr$_2$CrO$_4$
\cite{matsuno05}. From a conventional Drude-Lorentz fit we
estimate that the spectral weight of the free-carrier Drude
contribution is about 5\% of the total weight at 20 K.
Apparently, CaCrO$_3$ is very close to localized-itinerant
crossover. Its metallic behavior compared to insulating
Sr$_2$CrO$_4$ should be a consequence of the three-dimensional
crystal structure inducing larger band widths and thus smaller
$U_{\rm eff}$.

To further analyze the electronic structure, we carried out {\it
ab-initio} band structure calculations for the T=3.5\ K crystal
structure within the LSDA approximation using the linear
muffin-tin orbitals method \cite{Andersen-84}. Exchange constants
were computed from the total energies of different magnetic
solutions, using the crystal structure presented above. In LSDA,
CaCrO$_3$ is metallic in all studied magnetic structures: FM,
AFM-G (all nn spins antiparallel), AFM-A (AFM coupled FM $ab$
planes) and two AFM-C types with FM chains running in different
directions. In agreement with experiment, the AFM-C structure
with FM chains running along $c$ exhibits the lowest energy. The
calculated magnetic moment is $\mu = 1.52 \mu_B/Cr$, in good
agreement with the measured value of 1.2 $\mu_B$. The reduction
from 2$\mu_B$ expected for Cr$^{4+}$ ($S=1$) is caused by the
strong $pd$ hybridization.

Studying the exchange parameters allows one to understand the
apparently anisotropic magnetic structure. We find a strong AFM
interaction between nn spins within the $ab$ plane, $J$ = 80~K,
for the notation of magnetic interactions see Fig. 2a).
Surprisingly, also the nn exchange along $c$ is AFM and only
slightly smaller, $J'$ = 60~K, although the experiment finds FM
coupling in this direction. Its cause resides in a remarkably
strong AFM nnn interaction along the diagonal, $J''$ = 33~K.
Since $J'< 4J''$, the AFM $J'$ is overruled yielding the C-type
structure. Thus, the anisotropic magnetic structure develops due
to strong and anisotropic nnn interactions despite nearly
isotropic nn interactions. Also the diagonal nnn exchange within
the $ab$ planes is AFM, $J^{ab}_d$ = 22~K, but not sufficient to
overrule the nn $J$. The subtle balance of different interactions
may give rise to strong magnetoelastic coupling, explaining the
pronounced structural anomalies at $T_N$. The flattening of the
octahedron enhances the $d_{xy}$ occupation thereby increasing
$J$ and - more importantly - decreasing $J'$. Note that magnetic
interactions in the LSDA approach are due to the band magnetism
of itinerant electrons. Therefore, the rather large diagonal
coupling parameters are caused by strong $pd$ hybridization.

To check the importance of electronic correlations, we also
performed LSDA+U calculations \cite{Anisimov-91} with on-site
Coulomb interaction $U=3$~eV and Hund's rule coupling
$J_H=0.87$~eV \cite{Korotin-98}. Also in LSDA+U the ground state
is C-type AFM, but the electronic state is very different. In
LSDA+U, CaCrO$_3$ is an insulator with a gap of $E_g \sim$ 0.5~eV
(note that LSDA+U tends to overestimate $E_g$). In the LSDA+U
approach, the C-type magnetic structure is associated with orbital
ordering: one electron localizes in the $xy$ orbital at each Cr
site and provides the in-plane AFM interaction, the second
electron occupies alternating $1/\sqrt{2}(xz+yz)$ and
$1/\sqrt{2}(xz-yz)$ orbitals. According to the
Goodenough-Kanamori-Anderson rules this causes a FM interaction
along $c$. 
This state is very similar to the one reported for insulating
YVO$_3$ with G-type orbital order causing C-type magnetism
\cite{q1,q3}. We have searched for the orbital-order
superstructure reflections in CaCrO$_3$ by high-flux powder
neutron diffraction but did not find them although superstructure
reflections 10$^3$ times weaker than a strong fundamental
reflection would have been observed. Furthermore, a free
refinement of the orbital-order model with the high resolution
SPODI data does not yield any evidence for orbital ordering. We
may thus exclude an orbital order comparable to that in YVO$_3$
for CaCrO$_3$. LSDA+U evidently does not describe CaCrO$_3$
properly, but we nevertheless think that electronic correlations
are important in CaCrO$_3$ 
driving it close to a metal-insulator cross-over.

Summarizing our comprehensive investigation, combining
diffraction, macroscopic and optical studies, we conclude that
CaCrO$_3$ is a metallic and antiferromagnetic transition-metal
oxide. There are other metallic antiferromagnetic oxides known,
but these exhibit a reduced electronic and structural
dimensionality rendering CaCrO$_3$ unique. The anisotropic
$C$-type magnetic structure is explained by frustrating nnn
(diagonal) interactions. Apparently, the magnetic interactions in
\cacro ~ are governed by sizeable $pd$ hybridization, a generic
consequence of the high oxidation state associated with a small
or negative charge transfer gap.

\par This work was supported by the Deutsche Forschungsgemeinschaft
through SFB 608 and by the Dynasty Foundation and via projects
RFFI-07-02-00041, INTAS 05-109-4727, CRDF Y4-P-05-15 and
MK-1184.2007.2.

\end{document}